\documentclass{emulateapj}
\usepackage{graphicx} 
\usepackage{bm}
\newcommand{\bx}[0]{\bm{x}}
\newcommand{\bxp}[0]{\bx_+}
\newcommand{\bxm}[0]{\bx_-}
\newcommand{\by}[0]{\bm{y}}
\newcommand{\bu}[0]{\bm{u}}

\newcommand{\byc}[0]{\by_c}
 
\newcommand{\Ap}[0]{A_+}
\newcommand{\Am}[0]{A_-}

\begin{document} 
 
\title{Astrometric Perturbations in Substructure Lensing} 
\author{Jacqueline Chen\altaffilmark{1}, Eduardo Rozo\altaffilmark{2}, 
Neal Dalal\altaffilmark{3}, \&  James E. Taylor\altaffilmark{4}} 
 
\altaffiltext{1}{Argelander-Institut f\"{u}r Astronomie, Universit\"{a}t Bonn,
Auf dem H\"{u}gel 71,
D-53121 Bonn; 
       {\tt jchen@astro.uni-bonn.de}} 
\altaffiltext{2}{CCAPP Postdoctoral Fellow,
Department of Physics,
The Ohio State University, 1040 Physics Research Building
191 West Woodruff Avenue
Columbus, Ohio 43210-1117; 
       {\tt erozo@mps.ohio-state.edu}} 
\altaffiltext{3}{Canadian Institute for Theoretical Astrophysics, 
       University of Toronto, 
       60 St. George St., Toronto, Ontario, Canada M5S 3H8; 
       {\tt neal@cita.utoronto.edu}} 
\altaffiltext{4}{Department of Physics and Astronomy,  University of Waterloo
200 University Avenue West, Waterloo, Ontario, Canada N2L3G1;       {\tt  taylor@uwaterloo.ca}} 
 
\begin{abstract} 
 
In recent years, gravitational lensing has been used as a means to
detect substructure in galaxy-sized halos, via anomalous flux ratios
in quadruply-imaged lenses.  In addition to causing anomalous flux
ratios, substructure may also perturb the positions of lensed images
at observable levels.  In this paper, we numerically investigate the scale of such
astrometric perturbations using realistic models of substructure
distributions.  
Substructure distributions that project clumps near the Einstein
radius of the lens result in perturbations that are the least
degenerate with the best-fit smooth macromodel, with residuals at the milliarcsecond scale.  Degeneracies between
the  center of the lens potential and astrometric perturbations
suggest that milliarcsecond constraints on the center of  the lensing
potential boost the observed astrometric perturbations by an order of
magnitude compared to leaving the center of the lens as a free parameter.  In
addition, we  discuss  methods of substructure detection via
astrometric perturbations that  avoid full lens modeling in favor of
local image observables and also discuss modeling of systems with
luminous satellites to constrain the  masses of those satellites.
 
\end{abstract} 
 
\section{Introduction} 

The cold dark matter (CDM) paradigm predicts that the dark matter (DM) 
halos hosting most galaxies contain a large number of low-mass, 
compact subhalos within their virialized regions.  These subclumps, 
collectively referred to as substructure, may or may not contain 
luminous stellar tracers.  Since the observed number of dwarf 
galaxy satellites in the Local Group falls short of the expected 
number of subhalos by more than an order of magnitude 
\citep{klypin_etal99,moore_etal99}, then either the CDM model is 
incorrect or dwarf galaxies are biased tracers of DM in 
galaxy-scale halos \citep[e.g.,][]{spergel_steinhardt00,hannestad_scherrer00,hu_etal00,bullock_etal00,benson_etal02,somerville02,stoehr02,nagai_kravtsov05}.  If the CDM 
paradigm is correct, then the paucity of optical counterparts of 
substructures leaves 
us with few avenues for detecting and investigating 
DM substructure.  

One possibility for studying dark substructure around local galaxies is 
through its perturbative effect on kinematically cold systems  
like tidal streams \citep[e.g.,][]{mayer_etal02} or galactic disks  
\citep[e.g.,][]{toth_ostriker92,benson_etal04}. These tests probe only  
the closest and most massive substructures, and are only applicable  
in very nearby galaxies. As a result, this method provides little information  
to address fundamental questions such as the amplitude of the matter power  
spectrum on subgalactic scales or the detailed properties of dark matter  
itself.  
 
Another promising method of study is through the gravitational lensing
effects of substructure, referred to as substructure lensing.
Much of the previous work in this field has focused
on the effects of substructure on image magnifications and fluxes
\citep{mao_schneider98,metcalf_madau01,dalal_kochanek02,kochanek_dalal03,
rozo_etal06} or on time delays \citep{morgan_etal06}.  However, substructure  can perturb the deflection
angle, $\alpha$, of lensed images as well.  Image positions in lensed systems may not be subject to the same foregrounds  -- such as dust absorption --  as flux anomalies.
Further, since the
astrometric perturbation, $\delta\alpha$, is a steeper function of
subhalo mass than flux perturbations, it may provide a qualitatively
distinct probe of substructure properties.  

A study of astrometric perturbations is particularly timely given
the possibility of  
submilliarcsecond resolution observations of strong lenses.  For example, 
\citet{biggs_etal04} have presented observations of a four image jet 
source system, CLASS B0128+437, in which milliarcseond astrometric 
perturbations may have been detected.  
Previous studies of astrometric perturbations by substructures
have come to conflicting  conclusions about their overall size and
probability.    \citet{metcalf_madau01} use lensing simulations  of
random realizations of substructure in regions near images to  suggest
that in order to change image positions by a few tens of
milliarcseconds (mas), there must be subclumps with masses   $\gtrsim
10^8 M_{\sun}$  in Milky Way sized halos that are well aligned with
the images they perturb.  This alignment is likely to be rare in CDM,
although the probability would  increase in systems where the source
is elongated in a jet  \citep[][]{metcalf02}.  On the other
hand, \citet{chiba02} tests the size of astrometric  perturbations in
B1422+231 with a model of CDM subhalos as point masses   and finds
deflections of 10 to 20 mas, using subclumps with masses greater than
$\sim 2 \times 10^{8} h^{-1} M_{\sun}$.  Additional studies of
astrometric perturbations by single perturbers have focused on the
detection of substructure via distortion of a finite source
\citep{inoue_chiba03,inoue_chiba05b,inoue_chiba05}.  In addition,
\citet{pen_mao05} studied the effect of multiple lens planes on the 
rotation of lensed images.  
 
In this paper, we estimate the amplitude of astrometric perturbations
produced  by substructures using realistic substructure models  and
test the feasibility of  observing such perturbations by comparison
with the image positions given by a  best-fit lens macromodel.
While \citet{metcalf_madau01} suggest that single clumps need to be
well aligned  with the images in order to produce observable
perturbations, measurable effects may be possible in more general
scenarios, given the collective effects of entire substructure
distributions.  In addition, the kinds of perturbations in
these scenarios may be different.  For example, while the scenario of
a nearby subhalo that perturbs only a single image ensures that
degeneracies with best-fit smooth macromodels are small, large numbers
of more  distant substructure may affect multiple images and exhibit
large degeneracies  with the macromodel.  In addition to using
detailed lens modeling, we discuss  model-independent methods  for
identifying substructures, proposing a method using systems with
multiply  imaged jets.
 
Substructure lensing may also be used for comparisons between luminous
satellites and dark subhalos.  For example,  we can address one of the
open questions regarding CDM substructure:  why are some subhalos dark
and some luminous?  There have been many  proposed mechanisms to
explain which subhalos have stars and which  do not. One possibility is
that the efficiency of star  formation diminishes with decreasing halo
mass, perhaps due to  photoionization squelching
\citep{bullock_etal01}.  Another suggestion has  been that luminous
galaxies form only in the highest mass halos  ($M>10^9 M_\odot$), and
over time tidal stripping reduces subhalo  masses to the low values
($M\sim 10^8 M_\odot$) inferred for the  smallest local dwarfs
\citep{kravtsov_etal04a}.  Yet another possibility is  that the low
masses inferred for local dwarfs are in fact mistaken,  and instead
are much larger, $M\gtrsim 10^9 M_\odot$  \citep{stoehr_etal02,stoehr_etal03}.  
\citet{kazantzidis_etal04} find that subhalos do not experience the significant mass 
redistribution in their centers required to embed satellites in 
massive subhalos, and studies stellar kinematics in the dwarfs also dispute the existence of massive subhalos \citep[e.g.,][]{wilkinson_etal04}.   Here, we propose directly  measuring the masses associated with
luminous satellites projected near lensed  images, the results of
which would direcly test the \citet{stoehr_etal02} hypothesis.
 
The layout of the paper is as follows.  We begin by describing the
host halo and substructure models we will be employing in our analysis
in \S\ref{sec:models}.  The statistical properties of the astrometric
perturbations in our models are presented in  \S\ref{sec:obsres}, and
their degeneracy with macro-lens model parameters is discussed in
\S\ref{sec:modres}.  We discuss  model-independent, astrometric
signatures of substructure lensing in  \S\ref{sec:jets}.  Finally, the
effect of luminous substructures on astrometric  perturbations and
lens modeling is discussed in \S\ref{sec:lumsats}, and we present a
summary of  our results and conclusions  in section
\S\ref{sec:conclusions}.
 
 
\section{Numerical Estimates} 
\label{sec:models} 

In this section we present the host halo and substructure models we
used to analyze 
astrometric perturbations and describe
our methods of creating artificial lens systems with substructure, finding the image 
positions in such systems, and fitting for the best-fit smooth model.
 
\subsection{Halo Model} 
 
We begin by specifying the macromodel parameters of the lens system --
using  typical lens parameters -- and the background  cosmology used
to generate artificial lenses.  In particular, we choose a lens
redshift of   $z_l=0.5$, a source redshift of $z_s=2.0$, and a flat
$\Lambda$CDM cosmology with  $\Omega_m=0.3$, $\Omega_\Lambda=0.7$, and
$h=0.7$.  The halo  of the lensing galaxy is modeled as a singular
isothermal ellipsoid (SIE),   the projected density profile of which
is given by
\begin{equation} 
\kappa (\xi) = \frac{b}{2 \xi}, 
\end{equation} 
where $\xi$ is the projected radius, $\xi^2 = x^2 + y^2/q^2$ and
$q=0.9$ is our  fiducial value for the lens's axis ratio.  The
particular values for the ellipticity  and shear do not appear to be
significant to the results of this paper.  The length  scale $b$
corresponds  to the Einstein radius of the lens for circularly
symmetric profiles  ($q=1$). We take as our fiducial value
$b=1.05\arcsec$, corresponding to  the Einstein radius of a singular
isothermal sphere of mass $M=10^{13} M_\sun$   in our chosen cosmology
and source and lens redshifts.  Finally, we assume there is an
external tidal shear of $\gamma=0.16$  aligned with the major axis of
the halo ($\theta_\gamma =0$).  Our particular  choice $\gamma=0.16$
corresponds to the best-fit value for the external shear   obtained by
\citet{bradac_etal02} for the lens system B1422+231,  a system known
to exhibit anomalous flux ratios.
 
\subsection{Substructure Models}

We employ four different subhalo catalogs to populate our fiducial halo with
substructures.  The catalogs describe the three-dimensional position in the parent halo and the 
density 
profile of each subhalo.  In order to use these catalogs in our halo model, the 
substructure positions and density profiles must be projected along an axis.  The 
density profiles are projected by parameterizing the spherical density profiles in 
three dimensions, then using an approximate formula for the profile in two dimensions.  
The following section describes the substructure models in detail. 

For two of these models, we use substructure
catalogs from a DM cluster simulation with  $\Omega_{m} = 0.3,
~\Omega_{\Lambda} = 0.7, ~h=0.7, {\rm and} ~\sigma_{8} = 0.9$.  The
adaptive-refinement-tree (ART) code is used to run the simulation
\citep{kravtsov_etal97,kravtsov99}, using a $256^{3}$ uniform root grid 
covering the computational box of 80$h^{-1}$ Mpc and resimulating a 
cluster in the box with more particles and higher spatial resolution 
\citep{tasitsiomi_etal04}.  For this higher resolution simulation, 
the effective mass resolution is $m_{p}=3.95 \times 10^{7} h^{-1} M_{\sun}$ 
and the smallest cell size reached is $0.6 h^{-1}$ comoving kiloparsecs.  
Details of the simulations can be found in 
\citet{nagai_kravtsov05}.  A variant of the Bound Density
Maxima halo finding algorithm \citep{klypin_etal99b} is used to
identify subhalos.  Details of the algorithm used
to find subhalos can be found in \citet{kravtsov_etal04}.
 
\begin{figure}[h] 
\epsscale{1.} 
\plotone{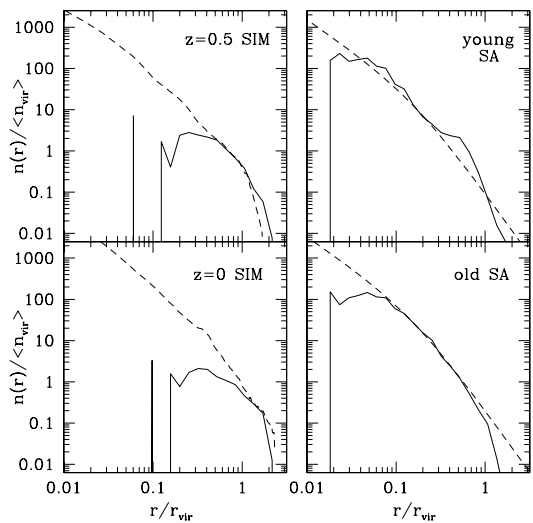} 
\caption{{\it Left:} The number density profile of subhalos from
numerical models   normalized to the number density within the three
dimensional virial radius,  $n(r)/\langle n_{\rm vir}\rangle$, for
$z$=0.5 (top) and $z$=0 (bottom).  The dashed line shows the dark
matter profile   normalized to the virial overdensity.  {\it Right:}
The number density profile of   subhalos from semi-analytic models
normalized to the number density within the   virial radius,
$n(r)/\langle n_{\rm vir}\rangle$, for a young halo (top) and an old
halo (bottom),   using all clumps with masses greater than $M = 10^{8}
M_{\sun}$.  The dashed line   shows a Moore profile with concentration
of 6.4, normalized to the scale radius.
\label{fig:nr}} 
\end{figure}

\begin{figure}[h] 
\epsscale{1.} 
\plotone{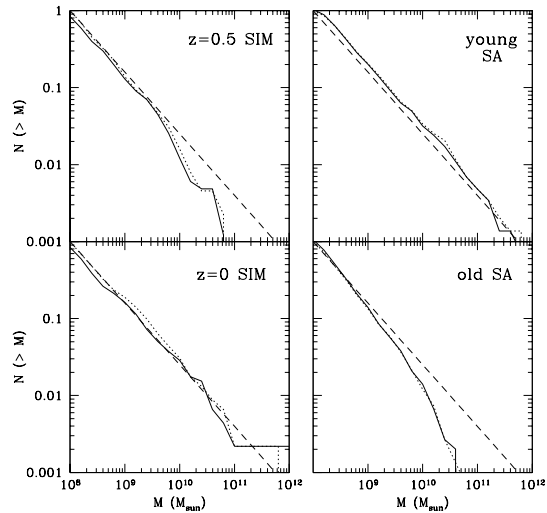} 
\caption{{\it Left:}  The cumulative mass function of subhalos from
numerical models   enclosed within the virial radius, for subhalos at
$z$=0.5 (top) and $z$=0 (bottom).  {\it Right:}  The cumulative
mass function of subhalos from semi-analytic models   enclosed within
the virial radius for subhalos for a young halo (top) and an  old halo
(bottom).  In all panels, the mass function from the three dimensional
mass profile is a dotted line, the mass   function derived from the
parameterization of the subhalos using Eqn. \ref{eqn:knew}   is a
solid line and the normalized $n \sim m^{-1.8}$ mass function is a
dashed line.
\label{fig:massf}} 
\end{figure}

We draw two different subhalo catalogs from the cluster simulation,
one at $z$=0.5, as the cluster is undergoing a major merger, and
another at $z$=0.  The mass of the parent halo itself is scaled down
to $10^{13} M_{\sun}$ and all subclumps are scaled
accordingly\footnote{Although the two catalogs are drawn from
different epochs, we use both for ad hoc substructure distributions in
the lens at the lens redshift and, therefore, fix the masses and virial 
radii to the same values.}.  Finally, clumps within the virial radius with 
mass $M \geq 10^8 M_\sun$  ($M/M_{vir} \geq 10^{-5}$) are chosen.  These
clumps correspond to   $N_{P} \geq 80$.  \citet{nagai_kravtsov05} show
that the mass function of the high resolution simulation converges
with that of a lower resolution simulation for $N_{P} \geq 80$.  We
fit the clumps with a Moore profile   \citep{moore_etal99},
\begin{equation} 
\rho(r) = \frac{\rho_{s}}{(r/r_{s})^{1.5}(1+(r/r_{s})^{1.5})} 
\end{equation} 
truncated at a radius $r_{t}$ using a Levenberg-Marquardt
minimization technique.  The   resulting overall mass fraction in
substructure is 10\% at $z$=0.5 and 13\% at $z$=0.    The left-hand
panels of Figures \ref{fig:nr} and \ref{fig:massf} show the number
density profile and cumulative mass function for both redshift
samples.  
 
 
Compared to galactic halo, the scaled cluster simulation may have a greater 
number of substructures \citep{zentner_etal05b} -- although in our substructure
catalogs, the substructure mass fraction does not seem significantly larger than 
expected at 10-20\%.  In addition, we may expect that the concentration of the subclumps  
in the cluster simulation may be smaller than in a comparable galactic halo;  the issue of
subhalo concentrations is
further discussed in Section \S\ref{sec:conc}.

The radial distribution of subhalos may be weakly 
dependent upon the mass scale of the parent halo \citep{gao_etal04,diemand_etal04}, 
with a cluster-scale simulation with fewer subclumps in the center of the halo than
a galaxy-scale simulation.  
In addition, simulations may suffer from  overmerging in the center of
halos, reducing the number of subclumps in the inner portions of the
halo  \citep{moore_etal96,klypin_etal99b}.  This effect may be
particularly  important in lensing studies, where substructure close
to the projected center of  the system produces the strongest
perturbations.  To get an independent estimate of how much
substructure  might survive in the inner parts of  galaxy halos, we
consider substructure distributions in two halos constructed using
the semi-analytic model described in
\citet{taylor_babul01,taylor_babul04};  these  models -- like our
simulation models -- do not include a galaxy component, but 
contain much more substructure in the central regions of the cluster
\citep{taylor_babul05},
thus likely providing an upper limit to the amount of substructure
in CDM halos.  
 
The semi-analytic models   assume spherical symmetry in the  input
orbits and the potential, and all clumps are parameterized by a Moore
profile,   truncated at $r_{t}$ and decreased in density by a fraction
$f_{t}$ as in \citet{hayashi_etal03} (equation 8):
\begin{equation}  
\rho(r) = \frac{f_{t}}{1 + (r/r_{t})^3}  
	\frac{\rho_s}{((r/r_{s})^{1.5})(1 + (r/r_{s})^{1.5})}. 
\end{equation} 
\citet{kazantzidis_etal04} suggests that the \citet{hayashi_etal03} relation may reflect lowered concentrations in the the subhalos simulation due to nonequilibrium initial conditions combined with numerical resolution rather than tidal shocks;  the issue of
subhalo concentrations is
further discussed in Section \S\ref{sec:conc}. 

The two semi-analytical substructure catalogs we choose correspond to
a dynamically  old and a dynamically young parent halo at $z$=0.  The
old halo has accreted   50\% of the parent halo's mass by $z$=3.2 and
90\% by $z=$0.67.  The corresponding  redshifts for the young halo are
$0.54$ and $0.04$.  Finally, the   old halo has a substructure mass
fraction of 10\%, while the young halo has a   mass fraction of 21\%.
We scale the host halo mass -- a galactic halo of 1.6$\times10^{12} M_{\sun}$ -- to $10^{13} M_\sun$  and scale
all the substructures accordingly, keeping only those substructures
with rescaled masses larger than $10^8 M_\sun$.  The right-hand
panels of Figures \ref{fig:nr} and \ref{fig:massf} show the resulting
number density profile and cumulative mass function for both samples.

In comparison to the simulation  profiles, the semi-analytic catalogs
have many more clumps in the center of the halo, while the number
density profiles of all the substructure models track the host halo dark 
matter profile at large
radii.  In addition, the mass functions of all the substructure models are similar,
$n \sim m^{-1.8}$, as expected from the results of high resolution
numerical   simulations
\citep{ghigna_etal00}.  In both the
simulation substructure models and the semi-analytic models, one of
the   distributions follows $n \sim m^{-1.8}$ to masses greater than
1\% of the   halo mass, while the other distribution falls off and has
fewer large-mass clumps.  We shall see that the presence or absence of
large mass clumps can have important  consequences on the distribution
of position perturbations.
 
In addition to specifying the general properties of the host halo and
its substructure population, lensing calculations require we specify
a line-of-sight projection axis.
We have chosen to project all halos along the major
axis of the host halo.  This choice for line-of-sight projection is
motivated by the fact that  it results in the most compact,  and
therefore most effective, lenses.  Consequently, one might expect most
observed strong lens systems to be projected along this axis.  Moreover, this
projection  leads to the largest substructure densities in the central
part of the halo,  which should maximize the impact substructures have
on the lensed  images \citep{zentner_etal05}.
 
As expected, the projected substructure distributions obtained from
the simulation  catalogs have fewer clumps in the central regions of
the  halo than the projected semi-analytic models. For instance,
within 2$\arcsec$ of the   center of the lens, the   $z=0.5$
simulation projection has no clumps and the $z=0$ simulation has 1
clump of   mass $3.8 \times 10^{8} M_{\sun}$, or $\sim$ 0.5\% of the
halo mass {\it that is enclosed within the Einstein radius}.  On the other hand,
both semi-analytic models have   7 clumps within this projected
radius, with total masses of $6.5\times 10^{9} M_{\sun}$ and   $7.1
\times 10^{9} M_{\sun}$ for the dynamically old and dynamically young
catalogs  respectively, or $\sim$ 5 - 10\% of the halo mass {\it that is enclosed
within the Einstein radius}.  For future reference, we note that the $z=0$
simulation model has the largest subclump of all the models, $\sim$
7\% of the halo mass, and this subclump is projected far from the Einstein 
radius of the host halo.
  
Rather than modeling the two dimensional mass  profile of each
individual substructure as a projected Moore or Hayashi  profile, we
use the approximation,
\begin{equation} 
\kappa (x)  = \frac{3.5 \kappa_{s}}{x^{1/2}+x^2} \label{eqn:knew}, 
\end{equation} 
where $\kappa_{s} = 2 \rho_{s} r_{s}$ for projected Moore profiles and  
$\kappa_s = 2 f_{t} \rho_{s} r_{s}$ for the Hayashi profile, and 
$x=2.4 R/r_s$ where $R$ is the projected radius.  
Our choice of profile is numerically motivated in 
that the deflection angle associated with our chosen profile has a simple  
closed form expression, whereas the projected Moore profile does not.  We  
show the mass functions of the substructure distributions using this  
approximation in Fig. \ref{fig:massf} (solid lines). Note our  
approximation does not appreciably bias the substructure mass functions.

\subsection{Generating and Modeling Artificial Lenses} 
 
Our host halo and substructure models are used to generate artificial
lenses as follows.  First,  the image positions of a given point source are
obtained by finding all roots  of the lens equation   using a
Newton-Raphson root finder with a gridded set of initial guesses.
Once the image positions of a source are obtained, we fit the
resulting lens system with a lens model that parameterizes the smooth
components of the lens only.  We then compare the true image
position with that obtained in the absence of substructures and
with that predicted by the best-fit model for each image in the lens. We repeat this
experiment to statistically sample the source plane in  order to
obtain distributions for the astrometric perturbations generated  by
the dark matter substructures.
 
The results of such a statistical comparison obviously depend on the
how the source plane is sampled.
While naively one might expect uniform source plane sampling
to be adequate, observed lens samples suffer from magnification
bias (brighter systems are more likely  to be observed than dimmer
systems) and consequently magnification weighted sampling 
is the most appropriate choice.
\citet{keeton_zabludoff04} show  that for sources with a power law
luminosity function, $dN/dS \propto S^{-\nu}$  with $\nu=2$,
magnification weighting in the source plane is equivalent  to uniform
sampling of the {\it image} plane.  The importance of this result
rests on the fact that  uniform image plane sampling is easy to
implement and the fortuitous fact that  the observed quasar luminosity
function in the largest lens surveys is roughly   a power law with an
index, $\nu \approx 2$. Consequently,   we have opted for
distributing sources along the source plane in accordance with uniform
image plane weighting in order to provide a closer match to
observations.
 
Best-fit models to artificial lenses are obtained by $\chi^2$
minimization  with a downhill simplex algorithm.   If convergence is
not achieved within a prescribed number of   steps, the original input
parameters (the macromodel   parameters) are perturbed and the
modeling is repeated until convergence is achieved.   The $\chi^2$
minimization puts priors on parity agreement and excessively magnified
images during modeling.  We eliminate extremely dim images, $\mu <
0.01$, and systems with excessively magnified images, $\mu > 50$.
Finally, an appropriate  best-fit model may not be found if the
simplex travels to an area of parameter space  that produces a number
of modeled images that is different from the observed number of
images.
 
The smooth models used in the  minimization are comprised of an SIE mass
distribution along with an  external shear component.  The model
parameters  are the Einstein radius $b$, the projected axis   ratio
$q$, the external shear $\gamma$, the orientation of the shear
$\theta_{\gamma}$, the orientation of the ellipticity $\theta_{q}$,
the   center of the potential $x_{\rm origin}$ and $y_{\rm origin}$,
and the   coordinates of the source position $x_{\rm source}$ and
$y_{\rm source}$.
 
Notice that lens systems with a single source position have nine
parameters to be fit  whereas there are only  eight observables in
quadruply imaged systems.  To make the system overconstrained, we
artificially hold the origin (i.e., the position of  the SIE) fixed at
$(x,y)=(0,0)$, thereby reducing the number of free parameters to
seven.  Observationally, this can be accomplished by fixing the center
of the mass   distribution to the observed position of the lensing
galaxy.  Recent work by  \citet{yoo_etal05} suggest that the position
of the lensing galaxy is an appropriate  approximation of the center
of the lensing potential.  For lens systems with more than a single
source position (i.e., jet sources), quadruply imaged   systems will
always be overconstrained without reducing the number of   parameters.
Note that in all sections, we restrict ourselves to four image
lenses.

\section{Intrinsic Astrometric Perturbations} 
\label{sec:obsres} 
 
In this section we investigate the intrinsic perturbations generated by  
substructure: the position difference between images
generated by a parent halo alone and images generated by a parent 
halo with substructure.  While observationally we are interested  
in the residuals relative to the best-fit model, comparisons to the 
intrinsic perturbations allow us to determine to what extent 
perturbations by substructure are degenerate with changes in the macromodel 
parameters of the best-fit lens.

Our initial expectations for intrinsic perturbations are colored by
our understanding  of single perturbers.  Consider the case of adding
a single perturber to a  smooth macromodel;  it is clear that the
astrometric perturbation will scale with the  the size and radial
position of the perturber, so that larger and more centrally located
subclumps produce larger perturbations.  In the case of substructure
distributions, however, it is unclear if  the most massive/most
centrally located substructures will dominate the astrometric
perturbations.  For instance, the steepness of   the mass function
means that there are many more small perturbers than large, so if the
former  act cooperatively they could in principle generate a large
perturbation.    Conversely, since oppositely positioned perturbers
generate equal and opposite  perturbations, the net effect of a large
number of substructures may cancel   out ensuring that rare, massive
substructures dominate the position perturbation of the images.
 
 
\begin{figure}[h] 
\epsscale{1.} 
\plotone{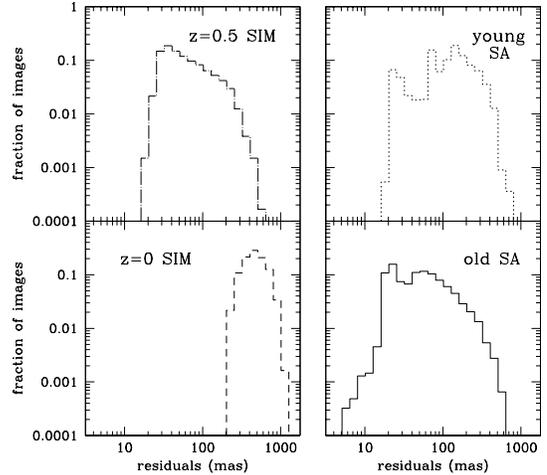} 
\caption{Histograms of the residuals between observed images in
milliarcseconds for   systems with substructure and without
substructure for quadruply-imaged single point   source systems. The
panels show the $z=0.5$ simulation distribution ({\it top-left})
dynamically young semi-analytic   substructure distribution ({\it
top-right}), the dynamically old semi-analytic   distribution ({\it
bottom-right}), and the $z=0$ simulation distribution  ({\it
bottom-left}). \label{fig:hist_real}}
\end{figure} 
 
 
Figure \ref{fig:hist_real} shows the intrinsic perturbation
distribution for  each of the fiducial
substructure models considered;  each system is quadruply-imaged and the residual
for each image position is calculated.    We note here that in this figure --
as in all of our subsequent figures involving histograms of residuals
-- shows  the results in a logarithmic scale.   Testing the entire
subclump distributions,  we find that the residual distributions all
have very   large peak perturbations, $\gtrsim 10$ mas.
Interestingly, our simulation-derived substructure models result in
residuals that are comparable to or larger than those of the
semi-analytic models, demonstrating that the intrinsic astrometric
perturbations are not necessarily dominated by nearby clumps.  In
addition, since the simulation models have extremely few or no
substructures near the Einstein radius of the lens and therefore the
perturbers must be located further away, we can infer that position
perturbations of different images in any lens configuration may be strongly
correlated.  A final interesting quality of our residual is that the
two simulations with the largest peak residuals are also the two
models that have very massive substructures.  This suggests that rare,
massive clumps may cause larger perturbations than the more abundant
smaller clumps.  However, even in the models where no such massive
substructures are present, the astrometric perturbations of the images
is still considerable.
 

\section{Modeled Residuals} 
\label{sec:modres} 
 
In general, the intrinsic perturbations just discussed are at least
partly   degenerate with macromodel parameters.  For instance,
substructure can change the  macromodel by adding to   the total
projected mass distribution and increasing the Einstein radius, $b$;
by changing the ellipticity   or orientation of the macromodel;  and
by adding external shear to the potential.  For example,  the
perturbations from a single subclump placed far from the lens are
degenerate with an external shear.   We expect then that the modeled
residuals, residuals between the observed image positions and the image 
positions 
predicted by the best fit smooth macro-model to the lens, will be
smaller than  the intrinsic residuals discussed above.
  
We test the extent to which the intrinsic perturbations  can be
accommodated with the macromodel and whether the remaining
perturbations (i.e.,  the perturbations relative to the best-fit model)
are large enough to be  detectable.  We consider only quadruply-imaged
systems and test two possible  lens modeling scenarios:  (1) a single
point source system modeled with the 7 parameters mentioned earlier
($b,q,\theta_q,\gamma,\theta_\gamma,\vec{r}_{source},$ with  the
center of the potential fixed) and (2) a jet source, approximated by
two source positions separated by 10 mas and modeled by 11 parameters,
including the center of the potential and two source
positions.   In jet source cases, we refer to each pair of images
produced  by the source positions together as an image of the jet, and
refer to the image of  each individual subsource as a subimage.

\subsection{Automated Lens Fitting}

 
\begin{figure}[h] 
\epsscale{1.} 
\plotone{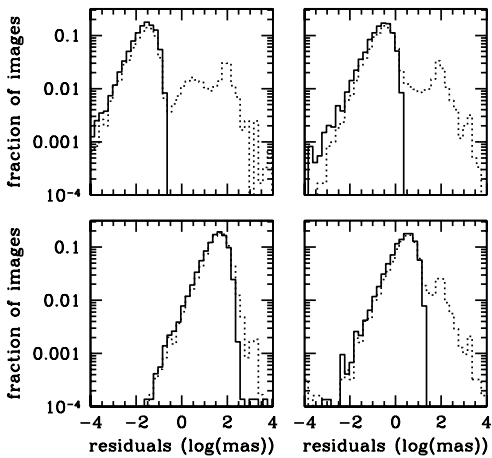} 
\caption{Histograms of the residuals between modeled and observed
images in   milliarcseconds for single point source systems using 7
fitted parameters and fixing   the center of the lens potential to
(x,y)=(0,0). Only four image systems are  plotted.   In each panel the solid line
indicates images with Gaussian errors and no substructure, while the dotted line is the 
same, where the
initial guess for  the Einstein radius is underestimated by 10\%.  Clockwise from the top-left panel,  Gaussian errors with a standard deviation of 0.1, 1, 10, and 100 mas are shown.  We
can see that a tail of large residuals is an artifact of our fitting
procedure, though the peak perturbation scale is robustly
determined. Consequently, in all future plots of residual
distributions we focus only on the scale at which the distribution
peaks.  The tail of high residuals comprises about $10\%$ of the total
number of sources considered for each model ($\approx$ 10000).  
\label{fig:hist_mod}} 
\end{figure} 
 

Before we move on to results, a discussion of one of the difficulties
in the automated lens fitting procedure is required.  The lens fitting
is done through a simple downhill simplex algorithm.  If our initial
guess for the best-fit model is not in the same $\chi^2$ valley as the
true best-fit model, our resulting formal best-fit model can be quite
far from the true best-fit smooth macromodel.  We illustrate  this
point in Figure \ref{fig:hist_mod}.  For this figure, we generated
lenses using the smooth host halo mass distribution only (no
substructure), and then added Gaussian perturbations of 0.1, 1, 10, and 100 mas to the 
image positions in successive panels.  The resulting lenses were then fit starting from two
different initial guesses for the best fit model.  For the first,
shown in Figure \ref{fig:hist_mod} as a solid line, we used the original mass
distribution used to generate the lens in the first place (i.e., the
"correct" smooth mass model).  For the second, shown with the dotted
line in Figure \ref{fig:hist_mod}, we simply lowered our initial best
guess for the Einstein radius by $10\%$.  In all cases, only four image 
systems are used and the residual from each image is calculated.  As is evident from Figure
\ref{fig:hist_mod}, in the first case the lens is always fit
correctly, and the residual distributions are sharply cut around the scale of the deviation, exactly as we would expect.  On the other hand, for the
second case we find that there is a large, unphysical tail of high
residuals. This tail was found in all of our  substructure runs, and
accounts for about $\approx 10\%$ of the systems fit.  This relation is true for Gaussian
perturbations regardless of the size of the perturbations.

In the
remaining sections of our work, we ignore this tail as an artifact of
our lens fitting algorithm.  More importantly, the {\it peak} of the
residual distribution is correctly recovered by our algorithm, and the
location of the peak in  our substructure runs has physical
significance.

\subsection{Substructure Degeneracies with Macromodels}

 
\begin{figure}[h] 
\epsscale{1.} 
\plotone{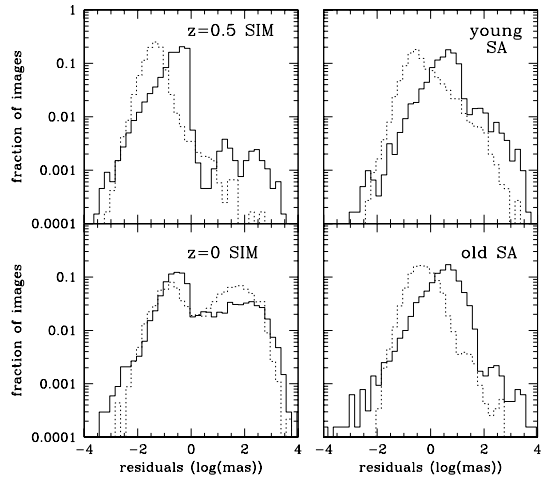} 
\caption{  Histograms of the residuals between modeled and observed
images in   milliarcseconds for single point source systems using 7
fitted parameters and fixing   the center of the lens potential to
(x,y)=(0,0) (solid) and for jet source systems using 11  fitted
parameters (dotted). Only four image systems are   plotted.  The
panels show the $z=0.5$ simulation distribution ({\it top-left})
dynamically  young semi-analytic   substructure distribution ({\it
top-right}), the dynamically old semi-analytic   distribution ({\it
bottom-right}), and the $z=0$ simulation distribution ({\it
bottom-left}).
\label{fig:hist_both}} 
\end{figure} 
 
 
We present the residual distribution for the single source and jet 
source scenarios using realistic substructure distributions in 
Fig. \ref{fig:hist_both}.  Here, the modeled residuals are small and, for the 
most part, significantly smaller than the intrinsic  residuals 
discussed in the previous section, with the peak perturbations 
occurring between .1 mas and 10 mas. The decrease in the size of the 
peak perturbations with respect to the intrinsic perturbations reflects 
the fact that there is significant degeneracy between the subclumps 
and the macromodel.   
 
Comparing the different substructure  distributions, we see that for
the single source case, the  semi-analytic substructure models produce
larger perturbations than the  simulation substructure models.  This
difference is evidence for the importance of having substructures
within the neighborhood of the Einstein radius of the lens in order
for the system to exhibit observable astrometric perturbations.  It is
also worth noting that the $z=0$ simulation substructure model
had the largest intrinsic position residuals and has the largest 
fraction of poor modeled fits -- demonstrating that a massive but distant 
clump changes the macromodel enough to be more difficult to find a best-fit 
model.  The acceptable modeled residuals, however, are comparable to that of 
the $z=0.5$ simulations, showing that those perturbations induced by a 
massive but distant clump are degenerate with an external shear. 
 
 
\begin{figure}[h] 
\epsscale{.6} 
\plotone{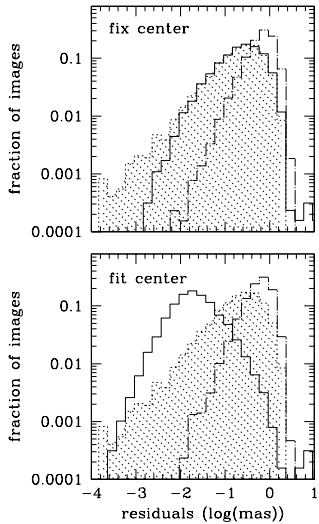} 
\caption{Histograms of the residuals between modeled and observed
images in   milliarcseconds for images with Gaussian errors with a
standard deviation of 1 mas and no substructure.  
For  jet sources, each subimage is
perturbed separately in the dot-dashed line but both subimages  are
perturbed by the same amount in solid line.  In both panels, the
shaded, dotted histogram  shows the distribution when employing single
point source systems, with the center of the  lensing potential fixed.
{\it Top:}  The center of the lensing potential is held fixed  for the
jet source systems.  {\it Bottom:}  The center of the lensing
potential is fit  for the jet source systems.
\label{fig:hist_nosub}} 
\end{figure} 
 
 
In addition to substructure model differences, there are significant 
differences between single point source systems, where we hold the center of 
the lens potential fixed, and jet source systems, in which the 
center is allowed to float.  When jet sources are used, modeled 
residuals are {\it smaller} than the corresponding residuals in the single
source case.  This result is contrary to our naive
expectations.  One could imagine taking each set of four subimages, 
and fitting each individually with a mass model. Since the best-fit 
models for each set of four subimages will in general differ, a 
single mass model for all eight images should result in larger 
residuals.  The resolution to this problem becomes apparent when we realize
that in fitting jet sources we have allowed the center of the potential 
to float, so the added freedom could result in a better fit.  Now, in 
practice, we find that this added freedom does result in better fits, 
but this didn't have to be the case, i.e., substructure 
position perturbations did not have to be degenerate with a change in 
the position of the lens's center {\it a priori.}
 
We illustrate this point in Figure  \ref{fig:hist_nosub}.  The figure
is produced by adding 1 mas Gaussian errors to images for  both
single point  source systems and jet source systems as lensed in the
absence of substructures (i.e., we use only the smooth halo component
to generate the artificial lenses).  In addition, we test two  methods
of adding the Gaussian errors to the jet source systems;  in one, each
subimage  is perturbed independently of the other, while, in the
other, each image is perturbed separately,  but the two subimages of
the each lensed image are perturbed in the same way.   We then proceed
to fit the artificial lenses.  As before, single point source systems
are  modeled by holding the center of the potential fixed.  Jet source
systems, on the other hand, are modeled both with the center of the
potential fixed and with the center of the potential allowed to float.
For the former case (top panel), we find that, just as we would
expect, the single source case (filled, dotted histogram) results in
smaller residuals than the jet source case (dot-dashed line) when each
subimage is perturbed independently.  When each
subimage pair is perturbed in the same way, the resulting histogram
(solid line) is essentially identical to that of the single point
source case, again in agreement with our expectations.  Turning now to
the bottom panel of Figure  \ref{fig:hist_nosub} in which the center
of the potential is allowed to float, we find that this added freedom
leaves the residual distribution of the jet source unchanged when each
subimage is perturbed independently.  When each member of a subimage
pair is perturbed in the same manner, however, the distribution of
residuals changes  dramatically, and in particular the peak residual
is an order of magnitude lower than what we found when the center of
the potential was held fixed.  Thus, our observed residual
distribution from the substructure lenses indicates that the position
perturbations generated by substructures are coherent on scales at
least as large as the assumed extent of our jet source.
 
 
\begin{figure}[h] 
\epsscale{1.} 
\plotone{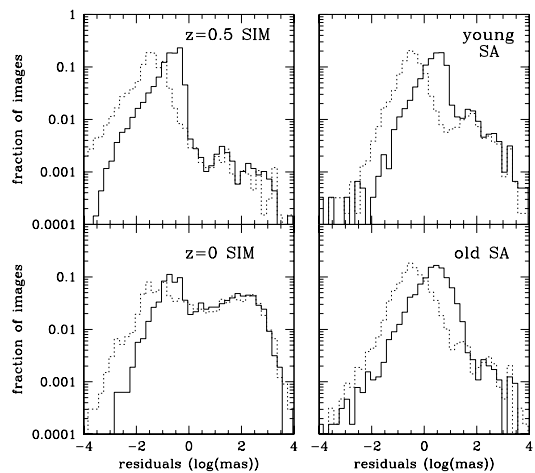} 
\caption{  Histograms of the residuals between modeled and observed
images in   milliarcseconds for single point source systems using 7
fitted parameters and Gaussian  priors on the center of the lens
potential of 5 mas (dotted) and 1 mas (solid).  Only  four image
systems are   plotted.  The panels show the $z=0.5$ simulation
distribution ({\it top-left}) dynamically  young semi-analytic
substructure distribution ({\it top-right}), the dynamically old
semi-analytic   distribution ({\it bottom-right}), and the $z=0$
simulation distribution ({\it bottom-left}).
\label{fig:hist_both_g}} 
\end{figure} 
 
 
\begin{figure}[h] 
\epsscale{.6} 
\plotone{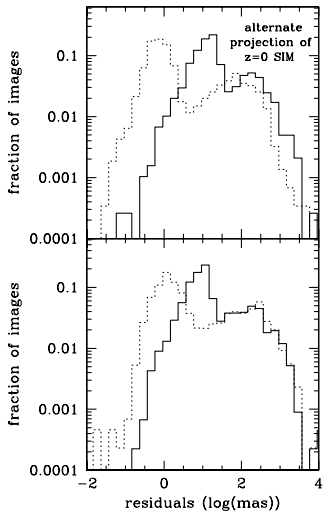} 
\caption{Histograms of the residuals between modeled and observed
images in   milliarcseconds for our alternate projection of the $z=0$
simulation substructure  model that projects a large subclump near the
center of the system. Only four image  systems are   plotted.  {\it
Top:}  Single point source systems using 7 fitted parameters and
fixing   the center of the lens potential are plotted in the solid
line and  jet source systems using 11 fitted parameters  are shown in
the dotted line.  {\it Bottom:}  Histograms of the residuals between
modeled and observed images in   milliarcseconds for single point
source systems using 7 fitted parameters and  Gaussian priors on the
center of the lens potential of 5 mas (dotted) and 1 mas  (solid).
\label{fig:hist_rot}}
\end{figure} 
 
 
\begin{figure}[h] 
\epsscale{1} 
\plotone{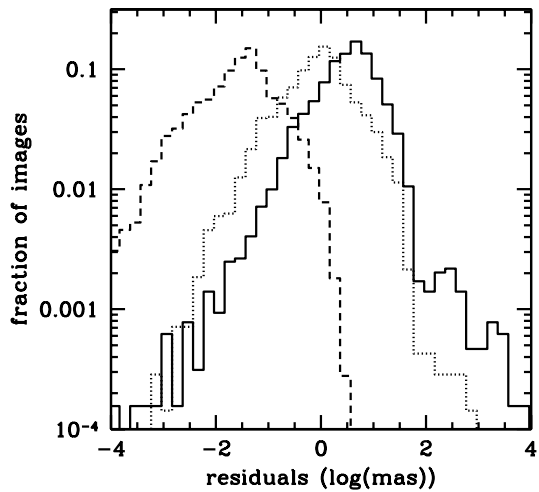} 
\caption{Histograms of the residuals between modeled and observed
images in  milliarcseconds for the dynamically old semi-analytic substructure model. Only four image  systems are   plotted.   Single point source systems using 7 fitted parameters and
fixing   the center of the lens potential are plotted.  The fiducial result is shown by a solid line.  Increasing the concentration by 2 (at fixed mass) is shown by a dotted line and increasing the concentration by 10 is shown by a dashed line. 
\label{fig:hist_conc}}
\end{figure} 
 
  
Given the importance of fixing the center of the potential,
an interesting question to ask is how well must the center of the
lensing potential be known for astrometric perturbations to be
sizeable.  We address this question in Figure \ref{fig:hist_both_g},
where we test the residual distributions of lenses with substructure
for single point source systems where the center of the potential is
constrained with a Gaussian prior.  As can be seen from the figure,  a
Gaussian prior of 5 mas does not result in much improvement  relative
to the case with no prior (the jet source case), while a prior of 1
mas  results  in a residual distribution similar to that obtained when
we fixed the center of the potential (our fiducial single point source
case).  Thus, we expect that the center of the potential must be known
to $\approx$ 1 mas accuracy for substructure perturbations to be
non-degenerate with an allowed shift of the center of the lensing
potential.  As discussed previously, the results of \citet{yoo_etal05} 
suggest that the assumption that the lensing galaxy represents the center 
of the lens potential holds to within 5 to 10 mas.  Detailed model 
fits of Hubble Space Telescope (HST) data shows that lens galaxy
astrometry down to 2 mas are achievable \citep[e.g.,][]{impey_etal98,lehar_etal00}.

\subsection{Massive Substructures Near the Einstein Radius}

One additional possibility we consider is what happens when an
extremely massive substructure projects near the Einstein radius of
the lens.  This question is relevant not only because one expect such
cases to result in the largest  residuals, but also because,
observationally, a large fraction of the current lens sample is seen
to have luminous satellites projected near the Einstein radius.
While luminous satellites can be directly included in the lens model,
it is still interesting to ask what residuals would a similar dark
substructure produce.

To address this question, we have chosen an an alternate projection of
the $z=0$  simulation model in which the most massive halo
substructure gets projected to within $\sim 2\arcsec$ of the halo
center.  We choose  the $z=0$ simulation model because 
this model contains the largest subclump among our four
substructure realizations with a mass of about $\sim$ 7\% of the {\it host
halo} mass.  A dark subhalo of this size is unlikely, but it is interesting
to test the extremes necessary for large astrometric perturbations.  

The resulting residual distributions for this substructure model are
shown in Figure \ref{fig:hist_rot} for both a single point source (top
panel, solid line) and a jet source (top panel, dotted line).  The
same basic qualitative features as seen in the fiducial cases are
found in this projection.  The peak residual for the single
source case is $\approx$ 10 mas and drops to $\approx$ 1 mas for the jet source case.  
The total amount of substructure within $2\arcsec$ of the lens center in this 
projection is significantly larger than those found for the fiducial substructure 
models, but the peak residuals are only somewhat larger.  
We also show in the bottom panel the residual distributions  obtained
using 1 mas (solid) and 5 mas (dotted) Gaussian priors on the center
of the lensing potential and using only a single point source.  We
again find that the center of the lensing potential must be known to
within about 1 mas for the position residuals to be sizeable.

\subsection{Dependence on the Concentration of Subhalos}
\label{sec:conc}

In the previous sections, we have preferentially chosen substructure realizations that would maximize the numbers of subhalos projected near the center of the lens -- using semi-analytic substructure models and choosing the line-of-sight along the major axis -- thereby maximizing the estimated signal from substructure.  For distributions with significant numbers of subhalos projected near the Einstein radius,  the peak of the residual distribution is a few milliarcseconds.  As discussed previously, however, the subhalos in  both the simulation-based substructure models and the semi-analytic substructure models could be biased to lower concentrations.  Higher concentrations may lead to larger deflections but may require better alignment with images.  In Figure \ref{fig:hist_conc}, the residual distribution for the  fiducial dynamically old semi-analytic substructure realization is compared to the same realization where the concentration of each subhalo is doubled while the mass and tidal radius of each subhalo is held fixed and to the realization where the concentration is increased by an order of magnitude.  Here, it can be seen that the peak of the residual distribution is {\it lowered} when the concentration of clumps is increased, so, in fact, our estimates likely represent the largest values for astrometric perturbations.  

\subsection{Halo-to-Halo Variation}

 
\begin{figure}[h] 
\epsscale{.45} 
\plotone{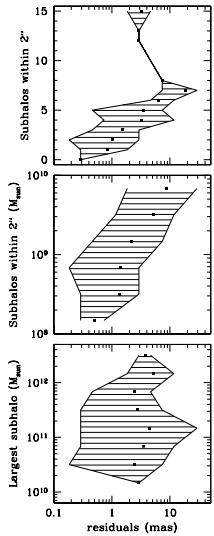} 
\caption{ The peak of the residual distributions for  4-image single point source lens systems for 48 semi-analytic substructure models using a random line-of-sight.  The average residual peak size is shown as square points, enclosed within the minimum and maximum peak residuals values.   {\it Top:}  The average peak residual compared to the number of subhalos within $2\arcsec$ of the halo center.   {\it Center:}    The average peak residual compared to the total mass in subhalos within $2\arcsec$ of the halo center.   {\it Bottom:}   The average peak residual compared to the mass of the largest subhalo in each distribution.  \label{fig:h2h}} 
\end{figure} 
 
 
 The fiducial substucture distributions show that the size of modeled residuals increases with a greater amount of subhalos projected near to image positions.  Halo-to-halo variation, however, could swamp this effect.  A full statistical analysis of halo-to-halo variation would require a more extensive understanding of both observational biases and substructure model parameters.  In this section, we test the possible range of the scatter using 48 different substructure distributions drawn from the semi-analytic models discussed previously and including the two fiducial semi-analytic models.  Testing single point source systems with a random line-of-sight, these semi-analytic substucture models have between zero and fifteen subhalos within 2 arcsecs of the center of the halo.  Most of the substructure realizations contain between three and five clumps within 2 arcsec of the center of the halo.  The number of models with a particular number of nearby clumps cannot be ascribed any particular significance, however, a relationship between the number of nearby clumps and the mean of the peak of the residual distribution may be seen in the top panel of Fig. \ref{fig:h2h}.  The range of residuals peaks in the bins is also shown in this figure, where, for the most populated bins, the scatter is significant, but not as large as the total range of peak distributions.  For all the models, the residuals fall between 0.1 mas and 30 mas.   A similar correlation is found between the average peak residual and the total mass of subclumps within 2 arcsec.  For comparison,  we also plot the average peak residual compared to the mass of the largest subhalo.  Here, no clear correlation can be observed.  
   
\subsection{Discussion}

Our results indicate that systems with massive substructures within
the area of the Einstein radius result in the largest residuals, of a few mas, which are an order of magnitude larger than residuals for substructure distribution with no nearby clumps.
Just as importantly, however, we have found that without milliarcsecond
constraints on the center of the lensing potential, astrometric
perturbations from substructures are sufficiently degenerate with
changes in the position of the lensing potential so as to reduce modeled residuals by an order of magnitude and making the residuals 
extremely difficult to detect on the basis of astrometric
perturbations alone.  \citet{yoo_etal05} 
have suggested that the assumption that the lensing galaxy represents the center 
of the lens potential holds to within 5 to 10 mas.   Astrometry of less than a few milliarcseconds for
lensed galaxies may be achieved using detailed model 
fits of Hubble Space Telescope (HST) data \citep[e.g.,][]{impey_etal98,lehar_etal00}.

Interestingly, \citet{biggs_etal04} have
presented a four image jet source system, CLASS B0128+437, in which
reproducing the observations with a smooth mass model appears to be
difficult, even while allowing the center of the lensing
potential float.  More specifically, the largest position residuals they
find are of order a few mas.  Such large perturbations are comparable to the
highest peak residuals we found, and suggest a large substructure mass
fraction at radii near the Einstein radius.  The errors in the original \citet{dalal_kochanek02} result are large enough 
to include both small and large substructure mass fractions (0.6\% to 7\% within the 90\%
confidence intervals).  In simulations, \citet{mao_etal04}  found that the substructure mass fraction at 
radii near the positions is small, $\sim$0.5\% -- as small as the amount of substructure found in our projected simulation halos which have small astrometric residuals.   In addition, studies of cusp relation anomalies with ray-tracing of simulations of galactic halos 
with substructure have implied that there in not enough substructure in simulations to account for the observed level of cusp anomalies  \citep{bradac_etal04,amara_etal06,maccio_etal06}

\citet{yoo_etal05} measured the displacement between the observed
position of various lensing galaxies and the position of the best fit
lensing model. Their finding of displacements between the two of order 
5 to 10 mas imply that the center of the lens potential and the position 
of the lensing galaxy are at least roughly coincident, but they also suggest 
that the displacements may indicate an alternative way to measure the 
substructure mass fraction.  
Displacements of this size are consistent with the
typical displacements we observed, modeling our clumpy lenses with 
smooth macromodels.  However, when we allowed the center of the 
potential to float using jet sources, we found that that even when there 
were no substructures or an extremely small substructure mass fraction 
within $2\arcsec$ of
the center of the lens, the center of the potential was still
displaced by a few to tens of milliarcseconds.  This suggests that
such displacements are sensitive to substructure far from the Einstein
radius of the lens, and hence interpretation of such observations as
limits on the substructure mass fraction in the central regions of
lenses might be subject to some caveats.  Further investigation of
whether the substructure mass fraction can be robustly estimated from
mis-alignment of the observed position of the lensing galaxy with the
best fit location for the center of the potential is clearly warranted.
 
 
\section{Detecting Substructure in Jet Systems} 
\label{sec:jets} 
 
Given the difficulties in using lens modeling to probe substructure 
distributions, unambiguous detections of substructure in individual 
lenses might be more robustly achieved through local measures
of astrometric perturbations.  
By offering more observational constraints, jet source 
systems seem particularly promising avenue of research.

\subsection{Astrometric Signatures of Substructure} 
 
Given a perfectly straight jet source small enough that
lensing can be locally linearized,  its images will also be perfectly
straight.  If one  were to observe a lensed jet with two images, one
straight and one with a kink, one might be tempted to conclude that
the kink must have been introduced by a local small scale
perturbation to  the lensing potential.  However, large kinks may also 
arise in the absence of substructures since small
deviations from linearity in the jet can be greatly amplified when the
lens mapping is nearly singular.\footnote{While bent radio
jets do not, in general, imply the presence of substructure,
some  lens configurations can be described that would unambiguously signal
the presence of localized perturbations.  For instance,  if two jets
in a fold configuration -- where two of four lensed images are close
together --  are bent in the same sense,  one would associate the
same parity to both images, violating a generic prediction of smooth
lensing potentials.  
}  In general, then, without
 {\it a priori} knowledge of the linearity of a 
jet source, some degree of modeling is required to detect substructure.

\subsection{Local Lens Modeling} 
 
In this subsection we consider one possible method for detecting
astrometric perturbations using purely local information from the
images (i.e., eschewing global lens modeling).  We consider lensed images
of sources where multiple source subcomponents are resolved and 
where each of the subcomponents is multiply imaged.  Moreover, we assume
the images are in a fold configuration, and focus our attention on
the close pair of images.  Assuming the
size of the source is smaller than any length scale associated
with the lensing potential, the mapping between the source and each of 
its images may be linearized.  Letting $A_+$ and $A_-$ denote
the inverse magnification tensors describing the mapping
from the positive and negative parity images, respectively,
onto the source, the two images must
themselves be related via a linear transformation,
 $\delta {\bf x_{+}} = U \cdot \delta {\bf x_{-}}$, where
  $U=(A_+)^{-1}A_-$ and $\delta {\bf x}_{\pm}$ denotes 
the image position vectors in any coordinate system chosen such that the origins in the lens plane maps to a single origin point in the source plane.  Moreover, the assumption that the images 
are in a fold configuration puts a constraint on the form of the
matrix $U$, so deviations on the form of the matrix $U$ from 
its expected structure may signal the presence of substructures
on scales comparable to or smaller than the separation between images.
The goal of this section is to investigate this possibility quantitatively.

Operationally, we proceed as follows.  Given a jet with three subimages, 
we define the best-fit linear transformation $U$ by minimizing the 
total residual $\langle \Delta r \rangle $ defined by
\begin{equation}
\langle \Delta r \rangle  = \sqrt{\frac{1}{N} 
	\left( \sum_{i=1}^N \left|(\bm{x}_+^{(i)}-\bm{x}_+^{(0)}) 
	- U \cdot
	(\bm{x}_-^{(i)}-\bm{x}_-^{(0)})\right|^2 \right )},
\label{eq:res}
\end{equation}
where the sum is over the position vectors, ${\bf x}_{\pm}^{(i)}$ is the 
image position of vector $i$, ${\bf x}_{\pm}^{(0)}$ is the image position 
chosen as the origin of the coordinate system, and $N$ is the number of 
terms in the sum.  For three jet subcomponents, $N=2$, so fitting
is overconstrained if the matrix $U$ has three free parameters or less.
Here, we constrain the form of the  
relative distortion matrix $U$ using the 
well known fact that for smooth lenses fold images are expected to
have equal and opposite magnifications as the image separation goes to zero.
We thus set the condition $\det(U)=-1$;  violations of this condition may be indicative of substructure.  It is worth noting that this 
constraint is coordinate independent, so  fitting can be done in 
any appropriate, observationally-defined coordinate system without any loss of generality.
For completeness, appendix \ref{app:expansion} explicitly performs a Taylor series
expansion
of the lensing potential around a specific point along the lens's critical
curve and in a specific coordinate system, though we found this more
detailed analysis to be {\it less} sensitive to substructure 
than the simpler approach presented here.

The properties of the distribution of the residual $\Delta r$ also depend
on the assumed properties of the jet source.  The magnitude of the residual
scales with distance between jet subcomponents, which we take to be $1$ mas
in our examples.  More interestingly, we will find that the distribution of
$\Delta r$ depends on the {\it jet bending angle}, the angle between the two
relative position vectors of the source subcomponents.  Thus, a zero degree
bending angle corresponds to a perfectly straight jet.

\begin{figure}[h] 
\epsscale{1.} 
\plotone{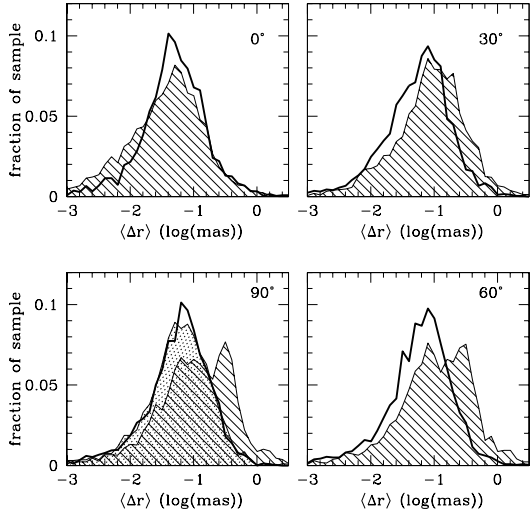} 
\caption{Distribution of $\langle \Delta r \rangle$ for different jet bending angles.  Clockwise from top-left, the bending angles are 0, 30, 60, and 90 degrees.  The thick solid line shows the distribution for the smooth, no substructure model, while the cross-hatched region shows the distribution for the old, semi-analytic substructure model in the 0, 30, 60 degree panels and increasing in deviation from the no substructure case  In the 90 degree panel, an additional shaded, dotted region shows the distribution for the $z=0$ simulation substructure model, which does not appreciably deviate from the no substructure case.}
\label{fig:method}
\end{figure}

Figure \ref{fig:method} shows the $\langle \Delta r \rangle $ 
distribution while enforcing $\det (U)=-1$ for various jet
bending angles.  Solid lines are obtained in the no substructure case,
and the cross-hatched histograms correspond to the case of the dynamically old semi-analytic substructure model.
We can see that for straight or nearly straight jets, the substructure
and no substructure cases are indistinguishable.  As the jet becomes
non-linear (the bending angle increases), the $\Delta r$ distributions
for the substructure case starts to extend to higher $\Delta r$ values,
implying that nonlinear jets may in principle detect substructure through
this local test, but only if the substructure density is high enough.  Indeed, 
the simulation substructure models are always indistinguishable from the
no substructure case, which is not unexpected given the lack of substructures
in the central regions of the halo.

 
Our result in Figure \ref{fig:method} might seem surprising: linear jets
are ineffective at detecting substructure without the full lens modeling.   As suggested strongly by section \S\ref{sec:modres}, the
simple picture of substructure producing
a bend in a single image while leaving the other untouched seems overly naive. 
Moreover, jets that are bent to begin with {\it can} detect substructure.
We interpret the bending angle dependence as follows: for a perfectly linear jet, 
observations of the jet constrain the lensing distortion along the single jet axis only, so
the correct amount of distortion can be reproduced in the substructure realizations
regardless of the $\det(U)=-1$ constraint.  For bent jets, however, the distortion 
along two different axis is probed, so {\it both} eigenvalues of the relative distortion
matrix can be constrained by observations. Since the $\det(U)=-1$ condition specifies
a relation between these two eigenvalues, it is not surprising that only when both
eigenvalues are resolved can the effects of substructures be discerned.  


\section{Luminous Satellites} 
\label{sec:lumsats} 
 
In the previous sections, we have focused on cases where the 
substructure producing astrometric perturbations is dark.  However, 
luminous satellites can perturb image positions as well, and in a number 
of strong lensing systems, bright satellites have been detected near 
the lensed images.  Such systems are important for several reasons. 
First, and most obviously, much more can be learned about the subhalo 
if its position is known: since the astrometric perturbation 
$\delta\alpha\propto m/r$ for a satellite of mass $m$ and distance 
$r$, a given perturbation can be produced by a nearby low-mass subhalo 
or a distant high-mass subhalo.  Additionally, by comparing the 
lensing properties of luminous satellites to those of dark subhalos, 
we can address why are some subhalos are dark and some are luminous.  Gravitational 
lensing can help resolve 
this question by directly 
measuring the masses associated with satellites falling near lensed 
images.  We illustrate this by considering two systems, MG J0414+0534 
\citep{trotter_etal00,ros_etal00} and MG J2016+112 \citep{koopmans_etal02}, which 
both have lenses at redshift $z\sim1$.  A key 
advantage of these systems over other lenses with known satellites 
(e.g. RXJ0911+0551) is that both have been observed with 
high-resolution VLBI imaging, giving astrometry at the 
sub-milliarcsecond level.  We infer the satellite masses in these 
systems by modeling the lensed image positions only, ignoring the 
image fluxes.  We hold fixed the position of the main lens galaxy and 
the satellite at the observed locations relative to the images, and 
model the mass distribution using a singular isothermal ellipsoid for 
the main lens and singular isothermal sphere for the satellite, also 
allowing for arbitrary external shear.  We run a grid of models for 
many values of the velocity dispersion of the satellite, optimizing 
over all other parameters, and define the likelihood as  
$L=\exp(-\chi^2/2)$.  For MG J0414+0534, astrometric data and 
uncertainties were taken from \citet{trotter_etal00}, while for MG 
J2016+112 these were taken from \citet{koopmans_etal02}.  The satellite in 
MG J0414+0534, referred to as Object X, has apparent magnitude 
$R=24.6$ corresponding to absolute magnitude of -19.4 for a 
$\Lambda$CDM cosmology with $\Omega_M=0.27$ and $h=0.71$.  The 
satellite G1 in MG J2016+112 has $I=24.6$, corresponding to an 
absolute magnitude of -19.6.   
 
\begin{figure}[h]
\plotone{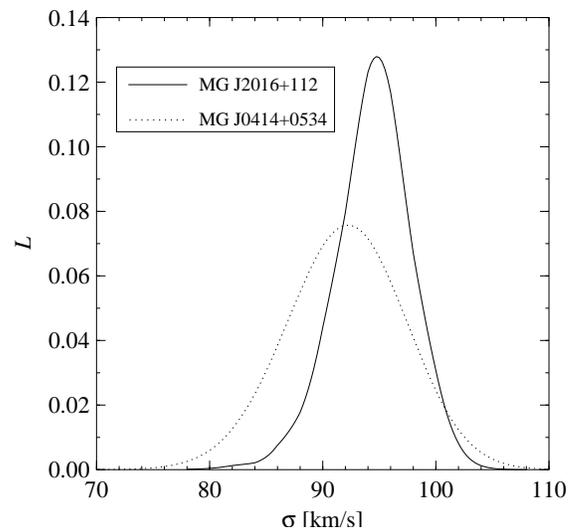} 
\caption{Lensing constraints on satellite masses.  The solid and 
  dashed curves, respectively, show likelihood as a function of 
  velocity dispersion $\sigma$ for the satellites in lens systems MG 
  J2016+112 and MG J0414+0534, where the satellite mass distribution 
  is modeled as a singular isothermal sphere.  
\label{fig:sig} 
} 
\end{figure}

The results of this procedure are plotted in Figure~\ref{fig:sig}.  For MG 
J0414+0534, the velocity dispersion $\sigma$ of the satellite Object X 
was found to be $81<\sigma<102$ km/s at 95\% confidence.  For MJ 
J2016+112, satellite G1 was found to have velocity dispersion 
$87<\sigma<101$ km/s at 95\% confidence.  Note that these correspond 
to quite large masses.  For comparison, the Milky Way Galaxy's largest 
satellite, the LMC, has an equivalent velocity dispersion 
$\sigma\approx v_{\rm rot}/\sqrt{2}\approx 50$km $\rm s^{-1}$ and a $V$-band 
integrated magnitude of $\sim -18$.  While the 
satellites in these systems are apparently rather massive, note that 
the lensing constraints were of sufficient quality to detect masses 
nearly an order of magnitude smaller. 
 
\section{Summary and Conclusions} 
\label{sec:conclusions} 
 
Astrometric perturbations from substructure lensing offers a new
avenue   for probing distributions of DM substructure in galactic
halos.  Previous   work in this field has mostly probed the effects of
single perturbers on   image positions
\citep{metcalf_madau01,inoue_chiba03,inoue_chiba05b,inoue_chiba05},
though \citet{chiba02} investigated the lens system B1422+231 using a
simple model in which CDM substructures were modeled as point masses
-- finding that typical astrometric perturbations are of order 10 to
20 mas.  Since,  observations of  the strong lens system B0128+437
with submilliarcsecond resolution have found astrometric residuals to
the best fit models to be slightly smaller at roughly 5 to 10 mas.
Such residuals, however, are large enough to be significant relative
to measurement errors \citep{biggs_etal04}.  In light of these
finding,  the question of how large astrometric perturbations by CDM
substructures  are is a very timely one.
    
In this paper, we use realistic models of substructure in order to
create populations  of artificial lens and estimate the residuals
between modeled and observed image positions.   Substructure
distributions were drawn from DM simulations and semi-analytic models
in  order to bracket the expected abundance of dark matter
substructures in the inner regions of galaxy size halos.  Our results
can be summarized as follows:
 
\begin{itemize} 
 
\item[1.]In general, astrometric perturbations from substructure are
partly degenerate  with the smooth macromodel.  Intrinsic image position
perturbations from substructure are typically of order $\gtrsim$ 10 mas,
even when few to no substructures are projected near the center of the
lens. Modeled residuals are significantly smaller by about an order of
magnitude.  
 
\item[2.]Substructure distributions in which clumps project near the
Einstein radius of the lens appear to be the least degenerate with the 
macromodel parameters and, therefore, cause the largest residuals between the
observed and modeled image positions.  When the center of the lens
potential is fixed, substructure models with significant amounts
substructure projected near the Einstein radius have modeled
residuals of order a few milliarcseconds.
 
\item[3.] Substructure astrometric perturbations appear to be coherent
on scales comparable to the extent of typical radio jets.  
This coherence significantly enhances the degree to which substructure
perturbations are degenerate with a change in the center of the
lensing potential.  Allowing the center of the potential to be fit as a 
parameter results in modeled residuals that are an order of magnitude 
smaller than would be the case if submilliarcsecond constraints of the 
center of the lensing potential were obtained.

\item[4.]  Tests of additional substructure realizations suggest that $\sim$10 mas 
may represent the largest residuals for any realistic substructure 
distributions where the center of the lens potential is fixed.  These substructure
realizations include nearby, massive dark subhalos and more concentrated halos.

\item[5.]  Local lens modeling of jet sources can in principle detect
substructure perturbations provided there exist substructures within
the Einstein radius of the lens in consideration and the jet sources 
are intrinsically bent.
 
\item[6.] Using gravitational lensing, it is possible to place constraints 
on the masses of luminous substructures of observed gravitational lenses 
through detailed lens modeling.  These constraints may be used to test 
proposed mechanisms for star formation in DM subhalos.

\end{itemize} 
 
 This work suggests that lens modeling may provide an avenue for detecting substructure through astrometric perturbations.  Substructure distributions which project subclumps near the Einstein radius of the lens halo cause the greatest perturbations;  however, substructure catalogs from simulations and semi-analytic models suggest that the scale of the perturbations will be small, on the order of a few milliarcsecond.  However, unlike anomalous flux ratios 
between lensed images which have been thoroughly investigated in previous works
\citep[e.g.,][]{mao_schneider98,metcalf_madau01,dalal_kochanek02,kochanek_dalal03,
rozo_etal06}, a complete accounting of the effects of astrometric
perturbations and  their observational implications has not yet been
done;  for example,  
observational biases remain unexplored.  
The results of future studies will offer a  better
understanding of submilliarcsecond resolution observations of strong
lens systems.
 
\acknowledgements 
 
This research was carried out at the University of Chicago, Kavli 
Institute for Cosmological Physics and was supported (in part) by grant 
NSF PHY-0114422. KICP is a NSF Physics Frontier Center.  J.C. was supported  
by the National Science Foundation (NSF) under 
grants No.  AST-0206216 and AST-0239759, and by NASA through grant 
NAG5-13274. Cosmological simulations used in this analysis were 
performed on the IBM RS/6000 SP4 system ({\tt copper}) at the National 
Center for Supercomputing Applications (NCSA). 
 
We are grateful to Stefan Gottl\"ober and Anatoly Klypin for providing us 
the high-resolution cluster simulation used in this study.

J.E.T was supported by the Leverhulme Trust, the U.K. Particle Physics and 
Astronomy 
Research Council (PPARC), the NSF under grant AST-0307859, and the DoE under 
contract DE-FG02-04ER41316. 
 
J.C. acknowledges the hospitality of the IAS, where part of this work was 
done. E.R.
thanks David Rusin for helpful discussions related to this work.

\bibliography{lens} 
 

\begin{appendix}

\section{Taylor Series Expression for the Relative Distortion Matrices of Folds}
\label{app:expansion}


\begin{figure}[h]
\epsscale{.8} 
\plotone{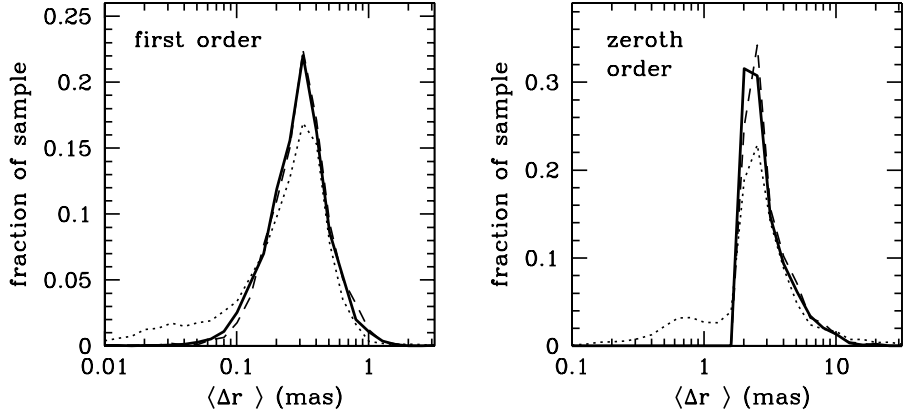} 
\caption{The $\langle \Delta r\rangle$ (${\rm mas}$) distribution for a 
sample of
realizations of fold image configurations with a 90 degree jet bending angle.  The thick, solid line represents
the no substructure case. The  dotted line indicates the 
dynamically old semi-analytic substructure distribution and 
the dashed line corresponds to the $z=0$ simulation
substructure catalog. {\it Left:} The first 
order expansion for $U$, Eqn. \ref{eq:U}, is tested. {\it Right:} The zeroth order 
expansion, Eqn. \ref{eq:U_zero}, is tested. }
\label{fig:expansion} 
\end{figure} 
 

In this appendix we present a brief derivation of the relative distortion matrix
$U$ for fold images.  We use the notation of \citet{schneider_book} throughout.
Let then $\bx$ and $\by$ represent position vectors on the lens 
and source plane respectively, and let $\by_0$ be the
point along the lens's caustic that is closest to the
source position $\bu$ of interest.  We choose the point $\by_0$
as the origin of the source plane, and its
image on the lens plane $\bx_0$ as the origin
of the lens plane. Moreover, we choose our axis
so that the inverse magnification tensor $A$ at $\bx_0$ is diagonal,
with the $(1,1)$ component non-zero. 
\\

With these assumptions, \citet[][equation 6.16]{schneider_book} show that 
one one possible parametrization
$\byc(\lambda)$ of the lens's caustic around $\by_0$ is given
by
\begin{equation}
\byc(\lambda) =\left( \begin{array}{c}
	\lambda \\
	\frac{bd-c^2}{2a^2d}\lambda^2 
	\end{array} \right).
\end{equation}
Note that $\lambda=0$ corresponds to $\by=0$, that is the
source plane origin.  
Given the source position of interest $\bu$, and using the
fact that, by definition, its distance to the caustic
$d^2(\lambda) = |\bu-\byc(\lambda)|^2$ is minimized at the
origin, we obtain the constraint equation
\begin{equation}
0 = \left.\left( \frac{d}{d\lambda}|\bu-\byc(\lambda)|^2\right)\right|_{\lambda=0}
\end{equation}
which simplifies to $u_1=0$.
In other words, the coordinate system we have chosen has the property
that the source position of interest $\bu$ falls exactly along
the $y$ axis.  Consequently, from now on we assume $\bu=(0,u)$.
\\

Given this source position, its corresponding images $\bxp$ and
$\bxm$ in the
lens plane are given by (see \citet{schneider_book} equation 6.17)
\begin{eqnarray}
(\bx_\pm)_1 & = & -\frac{cu}{ad} \\
(\bx_\pm)_2 & = & \pm\sqrt{\frac{2u}{d}}
\end{eqnarray}
where we have defined the Taylor series coefficients
$a=\phi_{11}^{(0)}$, $b=\phi_{112}^{(0)}$, $c=\phi_{122}^{(0)}$,
and $d=\phi_{222}^{(0)}$.  It is important to emphasize
that the above results implies that in the coordinate
system we have chosen, the image  separation
vector for fold images is along the $y$ axis. Moreover, since the $x$
axis is simply orthogonal to the $y$ axis, \it the coordinate system chosen
for the expansion can be observationally determined \rm up to an overall
displacement of the origin.

Returning to our derivation, in terms of the
expansion coefficients, the inverse magnification tensor $A(\bx)$ at
an arbitrary lens plane position $\bx$ near the fold 
is given by \citep[][equation 6.14]{schneider_book}
\begin{equation}
A(\bx) = \left( \begin{array}{cc}
	a+bx_2 & bx_1+cx_2 \\
	bx_1+cx_2 & cx_1+dx_2 
	\end{array} \right),
\end{equation}
so that the relative distortion matrix between the two images 
of interest is given by $U=(\Ap)^{-1}\Am$ where 
$A_\pm = A(\bx_\pm)$.  To lowest order in $u$, we find
\begin{equation}
U = \left( \begin{array}{cc}
	1+(\epsilon_1+\epsilon_2)	&	0 \\
	-k(1+\epsilon_2)	&  -1+(\epsilon_1-\epsilon_2)
	\end{array}\right)
\label{eq:U}
\end{equation}
where we have defined the quantities
\begin{eqnarray}
k & = & \frac{2c}{d} \\
\epsilon_1 & = & \frac{c^2-2bd}{\sqrt{2}ad^{3/2}}u^{1/2}\\
\epsilon_2 & = & \frac{3c^2-2bd}{\sqrt{2}ad^{3/2}}u^{1/2}.
\end{eqnarray}
Note that the inverse of this matrix is obtained by simply reversing the signs
of $\epsilon_1$ and $\epsilon_2$.  
The relative distortion matrix in the limit $u\rightarrow 0$ is
simply given by
\begin{equation}
U = \left( \begin{array}{cc}
	1 & 0 \\
	-k & -1 
	\end{array}\right).
\label{eq:U_zero}
\end{equation}
which has a single free parameter to be determined, and $\det(U)=-1$ as
it should.

Figure \ref{fig:expansion} shows the distribution of the average residual 
$\langle \Delta r \rangle$ given by equation \ref{eq:res} using the zeroth order and first
order expansions of the matrix $U$ computed above, using the same fold configurations as in 
Figure \ref{fig:method} for a jet bending angle of 90 degrees.  The coordinate system chosen for the expansion is approximated by setting the $y$ axis in the image plane to be in the direction of the vector between the fold images.  
As can be seen from the figure, the substructure and no substructure 
residual distributions are indistinguishable.

It is intriguing that a careful expansion of the lensing potential
resulted in a test that is \it less \rm sensitive to substructures than 
the more simple minded approach taken in the main body of this work.  It is
possible that this difference simply reflects the fact that the expectation
$\det(U)=-1$ is generic, and not particular to a specific point about which 
the potential is expanded.  Since the form of the 
matrix $U$ explicitly
depends on the point along about which the lensing potential is expanded,
it is possible that a formal expansion about a different origin from
the one considered here would result in a test of higher sensitivity to
substructures.
\end{appendix}

\end{document}